\documentclass[12pt,notoc]{JHEP3}

\usepackage{amsmath,amssymb,euscript,array,cite}

\input{epsf}
\usepackage{epsfig}

\def\N{{\cal N}}

\def\Tr{{\rm Tr}}

\def\det{{\rm det}}
\def\SU{\text{SU}}

\def\Dbarslash{\,\,{\raise.15ex\hbox{/}\mkern-12mu {\bar\D}}}
\def\Dslash{\,\,{\raise.15ex\hbox{/}\mkern-12mu \D}}
\def\delslash{\,\,{\raise.15ex\hbox{/}\mkern-9mu \partial}}
\def\delbarslash{\,\,{\raise.15ex\hbox{/}\mkern-9mu {\bar\partial}}}

\newcommand{\EQ}[1]{\begin{equation} #1 \end{equation}}

\newcommand{\SP}[1]{\begin{equation}\begin{split} #1
\end{split}\end{equation}}

\title{Higher rank Wilson loops from a matrix model}
\author{Sean A. Hartnoll${}^{a}$ and S. Prem Kumar${}^{b}$\\\\
${}^a$ DAMTP, Centre for Mathematical Sciences,\\
University of Cambridge,\\
Wilberforce Road,\\
Cambridge, CB3 0WA, UK.\\\\
${}^b$ Department of Physics,\\ University of Wales Swansea,\\
Singleton Park, \\
Swansea, SA2 8PP, UK.\\\\
\vspace{0.2 in}
\vspace{0.2 in}
E-mail: {\tt s.a.hartnoll@damtp.cam.ac.uk},
{\tt s.p.kumar@swan.ac.uk}}
\preprint{hep-th/0605027 \\ DAMTP-2006-32}
\abstract{We compute the circular Wilson loop of $\N=4$ SYM
theory at large N in the rank $k$ symmetric and antisymmetric tensor
representations. Using a quadratic Hermitian matrix model we obtain
expressions for all values of the 't Hooft coupling. At large and
small couplings we give explicit formulae and reproduce
supergravity results from both D3 and D5 branes within a systematic
framework.}

\begin{document}

\section{Background and motivation}

Wilson loops played a key role in the early development of the
AdS/CFT correspondence \cite{Maldacena:1997re}.
The identification of the Wilson loop in
${\mathcal N} = 4$ super Yang-Mills (SYM) theory with the dual fundamental
string \cite{Maldacena:1998im, Rey:1998ik} connected naturally with
ideas of understanding the `confining' strings of large $N$ gauge theory
as fundamental strings \cite{Polyakov:1997tj}. At finite
temperature, thermal Wilson loops enabled the identification of a
deconfinement transition in the theory \cite{Witten:1998zw} and allowed the
computation of the thermal interquark potential and screening length
\cite{Rey:1998bq, Brandhuber:1998bs, Landsteiner:1999up}.

There has recently been a renewed interest in Wilson loop
operators in the AdS/CFT correspondence, with an underlying theme of
understanding loops in higher representations of $SU(N)$.
This has seen a convergence of various different research directions.
Firstly, circular Wilson loops are half BPS operators and have been
conjectured to be computed
exactly, up to instanton corrections \cite{Bianchi:2002gz}, to all
orders in $N$ and $\lambda = g_{YM}^2 N$, from a quadratic Hermitian matrix model
\cite{Erickson:2000af, Drukker:2000rr}. This matrix model makes
nontrivial predictions for Wilson loops in higher representations
which constitute a test of the AdS/CFT correspondence incorporating
higher genus string effects \cite{Gross:1998gk}.

Recently, the matrix
model computation of a $k$ winding circular Wilson loop was
reproduced using a D3 brane embedded in $AdS_5$ carrying $k$ units of
worldvolume flux \cite{Drukker:2005kx}.
This beautiful match lead to further questions
as it was not clear which rank $k$ representation of $SU(N)$ the D3
brane was computing. A complete dictionary was later
proposed in \cite{Gomis:2006sb} relating supersymmetric Wilson
loops in an arbitrary representation to D3 and D5 branes in the dual
geometry. According to this dictionary, a single D3 brane with $k$
units of flux is dual to a
Wilson loop in the rank $k$ symmetric representation. It was therefore
an open question to explain why the $k$ winding computation of
\cite{Drukker:2005kx}, a priori certainly not the same as the rank $k$
symmetric respresentation, reproduced the supergravity result. This
confusion lead us to the study of general symmetric representations
in this paper.

A second line of interest has originated from the description of certain half
BPS operators of ${\mathcal N} = 4$ SYM theory carrying angular momentum
as `bubbling geometries' \cite{Lin:2004nb}. Wilson lines provide
another family of half BPS operators and so it is natural to search
for a bubbling geometry formalism describing the backreaction of the
D3 and D5 branes, as well as a field
theory understanding in terms of fermion droplets. Exciting progress
in this direction is currently being made \cite{Yamaguchi:2006te,
Lunin:2006xr, Okuyama:2006jc}. This line of interest inspired the
succesful matching of a bulk D5 brane computation
with a matrix model computation of the rank $k$ antisymmetric Wilson
loop \cite{Yamaguchi:2006tq}. However, that matrix model computation
required an assumption that is not a priori justified and does not
appear to indicate corrections to the supergravity result. This provided
the motivation for us to do an honest matrix model computation of the
antisymmetric representation in this paper.

Thirdly, computing the strong coupling thermal Wilson loops in higher
representations is crucial for mapping out the phase structure of
${\mathcal N} = 4$ SYM theory as a function of
't Hooft coupling and temperature \cite{Aharony:2003sx}. It is an
interesting open question whether there is a phase transition in the
high temperature phase between weak and strong coupling physics, with
implications for both black hole and plasma physics
\cite{Hartnoll:2005ju}. We recently initiated
the required study of D3 and D5 branes in the
AdS-Schwarzschild black hole background \cite{Hartnoll:2006hr}. The
results found for D5 branes in particular were curiously similar to
those for the BPS circular spatial loop at zero temperature
\cite{Yamaguchi:2006tq}.

There have also been recent works on single Wilson loops in the fundamental
representation \cite{Fidkowski:2004fc, Dymarsky:2006ve}.

In this paper we study circular Wilson loops in rank $k$ symmetric
and antisymmetric representations. Our computations will be in the
large $N$ limit of the Hermitian matrix model of
\cite{Erickson:2000af, Drukker:2000rr}. The expressions we derive will
be exact in $\lambda$; we will take the large $\lambda$ limit only
to obtain analytic results that may be compared with supergravity
computations. A key formula in our framework is the contour integral
representation of the symmetric and antisymmetric traces
(\ref{combin1}), which we then evaluate at large $N$ using a saddle
point approximation. We go on to reproduce the known supergravity
results for D3 and D5 branes \cite{Drukker:2005kx, Yamaguchi:2006tq}
as specific limits of this integral.

The results in this paper provide a check of the proposed holographic
dictionary for general representations \cite{Gomis:2006sb}, and furthermore
allows one to compute the corrections to the supergravity limit. This
last fact clarifies the regime of validity of the supergravity
computations. In the case of symmetric representations, we describe
an interesting phenomenon whereby the nature of the saddle point
changes qualitatively as a function of $k/N$ and $\lambda$. We finish
by deriving an explicit formula for the Wilson loops at weak coupling.

{\bf Note:} As we were typing up this work, the preprint
\cite{Okuyama:2006jc} appeared which has some overlap with our
results.

\section{Circular Wilson loops}

The circular loop in the fundamental representation was first computed by
Erickson, Semenoff and Zarembo \cite{Erickson:2000af} in perturbation
theory by resumming a specific class of planar rainbow and ladder
diagrams. It was further shown by these authors that cancellation of
coordinate dependence in the planar Wick contractions makes it
possible to map the result onto the solution of a large $N$ Hermitian
matrix model with quadratic potential. Subsequently this result was
argued by Drukker and Gross \cite{Drukker:2000rr} to be the exact
answer, up to nonperturbative instanton contributions
\cite{Bianchi:2002gz}, to all orders in $\lambda=g^2_{YM} N$ and
$1/N$.

In this section we calculate BPS circular Wilson loops in the
$k$-th rank antisymmetric and symmetric tensor representations
of the
$\SU(N)$ $\N=4$ theory. Specifically, we are interested in the expectation value of
the operator
\EQ{W_{{\cal R}}[U]\;=\;{1\over
    {\rm dim}\;[{\cal R}]} \;\Tr_{{\cal
      R}}\left[ {\cal P}
    \exp\left(i\oint_C ds(A_\mu {\dot x^\mu} + i\Phi_I
    \theta^I|\dot{x}|)\right)\right] \,,
\label{loop}
}
where ${\cal R}$ denotes the representation
and $\theta^I$
is a constant unit vector in ${\mathbb R}^6$. This choice breaks the
$SO(6)$ R-symmetry of the $\N=4$ theory down to $SO(5)$. The dimensions of
the rank $k$ symmetric and antisymmetric representions are
respectively $
{\rm dim}[S_k] = {(N+k-1)!\over k! (N-1)!}$ and ${\rm dim}[A_k]
={N!\over k! (N-k)!}$.
The contour $C$ will be
chosen to be a spatial circle at zero
temperature; this is then a BPS Wilson loop.

At finite temperature, the above operator with the contour $C$
chosen to be the Euclidean thermal circle yields the
Maldacena-Polyakov loop. This is known to be an order parameter for
the confinement/deconfinement transition in the strongly coupled
field theory \cite{Witten:1998zw}.
The Maldacena-Polyakov loop is not BPS in any natural
sense since the finite temperature theory is not supersymmetric.
Hence one does not expect the Maldacena-Polyakov loop to be related to the Hermitian
matrix integral which computes zero temperature circular BPS
Wilson loops. However, as mentioned in the introduction,
the works of \cite{Hartnoll:2006hr} and
\cite{Yamaguchi:2006tq} suggest a tantalising connection.
In particular, the operator \eqref{loop} was computed in
\cite{Hartnoll:2006hr} using a probe
D5 brane in the AdS-Schwarzschild black hole background yielding the
Maldacena-Polyakov loop for the finite temperature theory in the
antisymmetric tensor representation. Remarkably,
up to an
overall temperature dependent normalisation factor, the result
was found to be exactly the same as the zero temperature circular BPS
Wilson loop
in the antisymmetric tensor representation
\cite{Yamaguchi:2006tq}. This suggests that at least in the supergravity
$(\lambda=g^2_{YM}N\gg1)$ approximation there is a connection between
Maldacena-Polyakov loops and Hermitian matrix models. This paper,
however, will be concerned with the zero temperature theory and
supersymmetric loops.

\subsection{Circular Wilson loops from the matrix model}

The circular supersymmetric Wilson loop in the fundamental
representation of $\SU(N)$ was calculated in
\cite{Erickson:2000af} and \cite{Drukker:2000rr} and
shown to be computed by a Gaussian Hermitian matrix model.
The latter is defined via the partition function
\EQ{{\cal Z}=\int [d M]\; \exp\left(-{2 N\over \lambda} \Tr [M^2]\right)
\,,}
where $M$ is an $N\times N$ Hermitian matrix and $\lambda=g^2_{YM} N$
is the 't Hooft coupling of the $\N=4$ gauge theory.

Assuming the conjectured relation between BPS Wilson loops and the
matrix model for generic representations, circular Wilson loops in
any representation ${\cal R}$ of the $SU(N)$ gauge theory are then
given by
\EQ{{1\over {\rm dim}[{\cal R}]}\;\langle\Tr_{\cal R}[U]\rangle \;\;=\;\;
{1\over {\rm dim}[{\cal R}]}\;\langle\Tr_{\cal R} e^M\rangle_{mm}.}
The expectation value on the right hand side is computed in the
Gaussian matrix model above.
This is a remarkable conjecture which we
will employ to extract the Wilson loops in symmetric and antisymmetric tensor
representations.

In an eigenvalue basis where $M={\rm diag}\{m_1,m_2,
m_3,\ldots,m_N\}$, the expectation values of the operators in
the completely antisymmetric tensor representation can be expressed as
\EQ{\langle\Tr_{A_k}[e^M]\rangle_{mm} \;=\;
\sum_{1\leq i_1<i_2<\ldots<i_k\leq
  N}\langle \exp[m_{i_1}+m_{i_2}+\ldots+
m_{i_k}]\rangle_{mm} \equiv \langle \alpha_k \rangle_{mm} \,,} which
are simply $k$-th order symmetric polynomials ${\alpha}_k$ in the
eigenvalues $\{e^{m_i}\}$. The $\{\alpha_k\}$ are encapsulated in
the natural generating function for symmetric polynomials which is
the spectral curve or characteristic polynomial associated to the
matrix $e^M$
\EQ{F_A(t)\;\equiv \;\det(t+e^M) = \prod_{i=1}^N(t+e^{m_i}).}
It is easily seen that $\alpha_k$ is the coefficient of  $t^{N-k}$
in the spectral curve polynomial. It is most useful to express
this relation as a contour integral in the complex $t$ plane
\EQ{{\cal \alpha}_k= {1\over 2 \pi i}\oint dt \; {F_A(t)\over t^{N-k+1}} \,,
\label{residue}}
At any finite $N$ the contour may be centred either around the origin or
the point at infinity. Both definitions yield the same answer up to
an overall sign. For
infinite $N$ we will see that it is more convenient to think of the contour as encircling
the point at infinity.

The $k$-th rank symmetric representations are given by the
polynomials $\sigma_k$ in the eigenvalues $\{e^{m_i}\}$
\EQ{\langle\Tr_{S_k}[e^M]\rangle_{mm} \;=\;
\sum_{1\leq i_1\leq i_2\leq\ldots\leq i_k\leq
  N}\langle \exp[m_{i_1}+m_{i_2}+\ldots+
m_{i_k}]\rangle_{mm} \equiv \langle \sigma_k \rangle_{mm} \,.} The
generating functional for these polynomials is just the inverse of
the characteristic polynomial\footnote{We would like to thank Asad
  Naqvi for pointing this out to us.}. We find it convenient to
define it as
\EQ{F_S(t) \equiv {1\over {\rm det}(1-t e^M)}=
\prod_{i=1}^N {1\over (1-t e^{m_i})} \,.}
As in the antisymmetric case, we can extract the
polynomials $\sigma_k$ by a contour integral around the origin
\EQ{\sigma_k={1\over 2 \pi i}\oint dt \; {F_S(t)\over t^{k+1}} \,.
\label{residue1}}
Note that the generating function $F_S(t)$ decays as $t^{-N}$ for large
$|t|$, and has simple poles given by the positions of the eigenvalues of the
matrix $e^M$, namely at $t=e^{-m_i}$. Near the origin $F_S(t)\sim 1
+ \sigma_1 t + \sigma_2 t^2+\ldots$, and hence the $\sigma_k$ are
directly obtained from the above contour integral.

Our strategy will be to compute the expectation value of the
generating functions in the matrix model
\EQ{\langle F_{A,S}(t)\rangle_{mm}={1\over {\cal Z}}
\int \prod_{j=1}^N[d m_j]\; \Delta^2(m_k) F_{A,S}(t)\exp\left(-{2 N\over \lambda}\sum_{i=1}^N
m_i^2\right)\,,\label{eigen}} where, as usual we have written the
matrix integral in the eigenvalue basis at the expense of
introducing a Jacobian factor, namely the Vandermonde determinant
\EQ{\Delta^2=\prod_{1\leq i<j\leq N} (m_i-m_j)^2.}
In principle it should be possible to evaluate the above integral
for any finite $N$ and $\lambda$ by the method of orthogonal
polynomials and extract the expectation values of BPS Wilson
loops. However, in this paper we explore these expressions in the
limit of large $N$ and fixed $\lambda$, which is relatively
simpler and already very interesting.

\subsection{The large $N$ limit}

To take the large $N$ limit of the multi dimensional integral
\eqref{eigen}, we first rewrite it so as to make the $N$ dependence of
various terms explicit
\EQ{\langle F_{A,S}(t)\rangle_{mm}={1\over{\cal Z}}\int \prod_{i=1}^N [dm_i]\;
\exp(-S_{A,S}(\{m_i\}))\,,}
where
\EQ{S_{A,S}= {2 N\over \lambda} \sum_{i=1}^N m_i^2
  -\sum_{i<j}\log(m_i-m_j)^2 + \sum_{i=1}^N
  \left\{ \begin{array}{cc}
  -\log(t+e^{m_i}) & \text{for } S_A\\
  + \log(1 - t e^{m_i}) & \text{for } S_S
  \end{array} \right. \,,}
Clearly the tree level quadratic term is of order $N^2$, as is the
Vandermonde repulsive interaction between the eigenvalues. On the
other hand the insertion of the characteristic polynomial term in
the partition function only contributes at order $N$ in $S_{A,S}$.
Therefore, in the large $N$ limit, the integral will be dominated
by the usual Gaussian model saddle point configuration wherein the
eigenvalues condense onto a cut in the complex eigenvalue plane
with a spectral density characterised by the Wigner semi-circle
distribution (see e.g. \cite{Dijkgraaf:2002fc} and references
therein)
\EQ{\rho(m)={2\over \pi\lambda}\sqrt{\lambda-m^2}\,,\qquad\quad
-\sqrt\lambda\leq m\leq \sqrt\lambda.}
The spectral density is normalised so that
\EQ{\int_{-\sqrt\lambda}^{\sqrt\lambda}dm\;\rho(m)=1\,.}
The large $N$ expectation values of the generating functionals are
thus
\begin{eqnarray}
\langle F_{A}(t)\rangle_{mm} & = & \langle
  \exp\left({\Tr\log[t+e^M]}\right)\rangle_{mm} \nonumber \\
 & \rightarrow & \exp\left[ N \int_{-\sqrt\lambda}^{\sqrt\lambda}
dx\;\rho(x)\log(t+ e^x)\right]\label{largen} \,,
\end{eqnarray}
and
\begin{eqnarray}
\langle F_{S}(t)\rangle_{mm} & = & \langle
  \exp\left({-\Tr\log[1-t\;e^M]}\right)\rangle_{mm} \nonumber \\
 & \rightarrow & \exp\left[ -N \int_{-\sqrt\lambda}^{\sqrt\lambda}
dx\;\rho(x)\log(1 - t e^x)\right]\label{largen1} \,.
\end{eqnarray}

At this point we note that in the strict large $N$ limit there is
a drastic change in the analytic structure of the characteristic
polynomial $\langle F_A(t)\rangle_{mm}$. This can be understood in
two different ways. Firstly, for any finite $N$ the representation
of the polynomial as $\exp(\Tr\log(t+e^M))$ yields only an $N$-th
order polynomial upon a formal power series expansion around the
origin. The coefficients of higher powers of $t$ cancel due to
trace relations for large powers of the $N\times N$ matrix $e^M$.
At infinite $N$, as is evident from \eqref{largen}, the power
series in $t$ around the origin has infinitely many terms. This is
due to the fact that the $N\rightarrow\infty$ limit has already
been taken and one does not see traces involving powers of $e^M$
that are larger than $N$.

In addition, we note that on the real axis for $t <
-e^{-\sqrt\lambda}$ the expression develops an imaginary part.
This can be traced to a finite branch cut
discontinuity along the negative real axis. The discrete zeros of
the characteristic polynomial, which are the eigenvalues of $e^{M}$,
coalesce into a continuum in the large $N$ limit producing the
branch cut along $-e^{\sqrt\lambda}\leq t\leq
-e^{-\sqrt\lambda}$. Hence at infinite $N$ our expression for the spectral
curve does not yield a simple polynomial since it has a branch cut
singularity in the $t$ plane along with a pole of order $N$ at
infinity.

The generating function $\langle F_S(t)\rangle_{mm}$ for the symmetric
Wilson loops also 
has a branch cut on the real axis for $e^{-\sqrt\lambda}\leq t\leq
e^{\sqrt\lambda}$, which is due to the coalescing of $N$ simple poles in
the large $N$ limit.

The large $N$ circular Wilson loop in the rank $k$ antisymmetric
representation is then
\EQ{{\langle\Tr_{A_k}[U]\rangle\over {\rm dim}[A_k]}={{k!(N-k)!\over
      N!}}\oint_C 
  {dt\over 2\pi i}\;{t^{k-1}}\;
\exp\left[
    N \int_{-\sqrt\lambda}^{\sqrt\lambda}
dx\;\rho(x)\log(1+ e^x\; t^{-1})\right] \,, \label{integ}} where
we have absorbed the factor of $t^N$, see \eqref{residue}, into
the exponent and the contour of integration $C$ is to be specified
below. For the rank $k$ symmetric loops at large $N$, we have
similarly
\EQ{{\langle\Tr_{S_k}[U]\rangle\over {\rm dim}[S_k]}={{k!(N-1)!\over (N+k-1)!}}\oint_C
  {dt\over 2\pi i}\;{1\over t^{k+1}}\;
\exp\left[-
    N \int_{-\sqrt\lambda}^{\sqrt\lambda}
dx\;\rho(x)\log(1-e^x\; t)\right]\,.\label{integ1}}

The problem of computing the Wilson loops is reduced to
computing residues. This is quite straightforward in
principle for finite $k=1,2,\ldots$ A more interesting question is
whether it provides a useful way to explore the behaviour of these loops for
generic values of $k$ that are comparable to $N$.

\subsection{Wilson loops at $k\sim N$}

A compact expression for the rank $k$ Wilson loops is obtained
after a $t\rightarrow1/t$ variable change in \eqref{integ} so that
for the
 symmetric (S) and antisymmetric (A) loops we have
\EQ{\langle\Tr_{S_k,A_k}[U]\rangle={1\over 2\pi i}\oint_{C}
  dt\;{1\over t^{k+1}}\;
\exp\left[\mp
    N \int_{-\sqrt\lambda}^{\sqrt\lambda}
dx\;\rho(x)\log(1\mp e^x\; t)\right]\,,\label{combin}} where the
$-$ sign corresponds to the symmetric representation. The contour
$C$ is chosen around $t=0$ and this encapsulates the prescription
that the $k$-Wilson loops are the coefficients of positive powers
of $t$ in the Laurent series around $t=0$. Indeed, this
prescription works as we obtain
\EQ{\langle\Tr_{S_0,A_0}[U]\rangle=1 \,, \label{one}}
\EQ{\langle\Tr_{S_1,A_1}[U]\rangle=N\int_{-\sqrt\lambda}^{\sqrt\lambda} dx
  \rho(x) e^{x}= \langle \Tr[U]\rangle \,,}
\EQ{\langle\Tr_{S_2,A_2}[U]\rangle={1\over
    2}\langle\Tr U\rangle^2
\pm{1\over 2} N\int_{-\sqrt\lambda}^{\sqrt\lambda} dx
  \rho(x)
  e^{2x}={1\over 2}\left(\langle\Tr U\rangle^2\pm\langle\Tr
  U^2\rangle\right)\,. \label{three}}
This is not surprising but provides a simple check of the
prescription. Note that strictly speaking, the expressions (\ref{one}) to (\ref{three})
are not quite reliable within the large $N$ limit. We have kept
terms which scale differently with $N$, while replacing
$\langle(\Tr U)^n\rangle$ with $\langle\Tr U\rangle^n$ which is of
course a consequence of the $N\rightarrow \infty$ limit. In any
case, the direct approach of evaluating explicitly the residue at
the origin is not useful for computing Wilson loops with $k$
comparable to $N$. To access these we need to evaluate the saddle
point of the integrand at large $N$.

\subsection{Large $N$ saddle point}

The integral \eqref{combin} yields the Wilson loops in the large
$N$ theory and therefore only strictly makes sense in that limit.
In the $N\rightarrow\infty$ limit we only need to consider the
contribution from the saddle points of the integrand.

We are interested in the regime $k\rightarrow\infty$ with $k/N$
fixed. We introduce the fraction
\EQ{f= \frac{k}{N} \,,}
and treat it as a continuous variable in the large $N$ limit. In
addition, it is natural to perform the variable change
\EQ{t=e^{\sqrt\lambda z}\,,}
which takes us from the complex plane to the cylinder. The points
$t=0$ and $t=\infty$ are mapped to the points $z=\mp \infty$
respectively. 

For the antisymmetric case the branch cut singularity lying along the
interval $(-e^{\sqrt\lambda},-e^{-\sqrt\lambda})$ in the $t$ plane is
mapped to the line joining $z= -1+i{\pi\over\sqrt\lambda}$ and $z=
1+i{\pi\over\sqrt\lambda}$ and images thereof under translations
by multiples of ${2\pi
  \over{\sqrt\lambda}}i$ in the $z$ plane.  In the symmetric case the
branch cut along $e^{-\sqrt\lambda}< t< e^{\sqrt\lambda}$ is mapped to
the interval $-1< z< +1$ along the real axis and its images in the $z$ plane.

Using the explicit form of
the Wigner semi-circle distribution and appropriately rescaling
variables we find from (\ref{combin})
\EQ{\langle\Tr_{S_k,A_k}[U]\rangle={{\sqrt\lambda}\over 2\pi i}\int_{C}
  dz\;
\exp\left[\mp N \left(
    {2\over\pi}  \int_{-1}^{1}
dx\;{\sqrt{1-x^2}}\log(1\mp e^{-{\sqrt\lambda}(x-z)})\pm
f\sqrt\lambda z
\right)\right].\label{combin1}}
The contour $C$ lies to the left of the branch cut, in the region
${\rm Re}\; z< -1$ and winds once around the $z$ plane cylinder.

In the $N\rightarrow\infty$ limit, we expect the integral to be 
dominated by
the saddle points of the exponent. Hence we will first look for these.
The saddle point equations are
\EQ{{2\over \pi} \int_{-1}^{+1} dx {\sqrt{1-x^2}\over{1\mp
      e^{\sqrt\lambda(x-z)}}}\pm f = 0 \,.\label{saddle}}
The choice of upper signs yields the symmetric case.
These equations will in general have solutions in the complex $z$
plane. Note that for certain values of $z$ we need to worry about
potential divergences in the integrand in (\ref{saddle}). The integral
is then given by its Cauchy principal value along with an imaginary part. 
Such situations  
occur precisely when the solution lies on the branch cut and is not 
a stationary phase point.
An explicit
analysis of these equations is hard for generic $\lambda$. Instead we first
study the limit of large $\lambda$ which can be directly compared
with supergravity calculations.

\subsection{Supergravity limit}

The supergravity limit, $\lambda\rightarrow\infty$, leads to a
drastic simplification of the exponent in \eqref{combin1} and the
resulting saddle point equations \eqref{saddle}. For the rest of this
section, we will
treat the antisymmetric and symmetric cases separately.

\underline{\bf Antisymmetric}: This corresponds to the equation
\eqref{saddle} with the lower sign choice. Separating out the saddle point
equation into real and imaginary pieces, we find that the imaginary
part of the equation leads to the condition
\EQ{\sin\left({\sqrt\lambda}\;{\rm Im}\; z\right)\;=\;0.}
While this has the general solution ${\sqrt\lambda}\;{\rm Im}\;
z=\;n\pi$, we will see below that only the solution
${\rm Im}\;z =0$ (and images) is picked out as the
saddle point.

Now we turn to the real part of the saddle point equation. We first 
divide the $z$ plane into three regions: ${\rm Re}\; z<-1$, $-1\leq {\rm Re}
\; z\leq 1$ and ${\rm Re}\; z>1$. In each of these three regions the
limit of large $\lambda$ leads to distinct asymptotic expressions.
The saddle point equation 
can be expanded in powers of
$e^{-\sqrt\lambda}$ in the large $\lambda$ regime to yield
\EQ{ -f \;+ \;O(e^{-{\sqrt\lambda} (|{\rm Re}\;z|-1)})=0\qquad 
[{\rm Re}\;z <-1] \,,}
\EQ{{{2}\over \pi}\int_{-1}^{{\rm Re}\;z}dx
  {\sqrt{1-x^2}} -  f\;+ O(e^{-{\sqrt\lambda} (|{\rm Re}\;z|-1)})=0\qquad 
[-1\leq {\rm Re}\;z \leq 1]\,,\label{sad}} 
and finally
\EQ{(1- f)\;+ O(e^{-{\sqrt\lambda} (|{\rm Re}\;z|-1)})=0\qquad [{\rm Re}\;z > 1] \,.}
It is clear that in the regions ${\rm Re}\;z > 1$ and  
${\rm Re}\;z < 1$ there exist no consistent solutions in the strict
$\lambda\rightarrow\infty$ limit for fixed $f$. 

On the other hand, in the region $-1\leq {\rm Re}\; z\leq 1$ we have a
nontrivial equation that in general has solutions. Recalling that the
branch cut extends between $z=-1+i{\pi\over\sqrt\lambda}$ and
$z=1+i{\pi\over\sqrt\lambda}$, we conclude that the stationary phase
point lies at ${\rm Im}\; z=0$, on the real axis at
\EQ{z=\tilde z \in {\mathbb{R}} \,,}
where $\tilde z$ satisfies
\EQ{\cos^{-1}\tilde z-\tilde z\sqrt{1-\tilde z^2}=\;\pi(1-f)\,,}
It is natural to rewrite this in an angular parametrisation
\EQ{\tilde z= \cos\theta_k \,,}
where the subscript $k$ associates the solution to the rank $k$
representation. This gives
\EQ{\pi (1-f) =(\theta_k-\sin\theta_k\cos\theta_k) \,.\label{asol}}
Note that under $k\rightarrow N-k$, or equivalently $f\rightarrow 1-f$,
the angular variable $\theta_k$ transforms to $\pi-\theta_k$. We will
see in a moment that under this transformation the integral (\ref{combin1}) remains
invariant, reflecting the fact that antisymmetric representations of
rank $k$ and $N-k$ are in fact identical.

The saddle point value of the integral is determined by the exponent
in (\ref{combin1}). Putting $z = \tilde z$ and taking the large
$\lambda$ limit of (\ref{combin1}) we obtain
that the antisymmetric Wilson loop at large $k\sim N$ evaluates
to
\EQ{W_{A_k}[U] = {1\over {\rm dim}[A_k]}\langle\Tr_{A_k}[U]\rangle= \exp\left[{2 N\over 3\pi}
\sqrt\lambda\sin^3\theta_k\right] \,.\label{asym}}
Here we have only kept the leading terms in the strict large $\lambda$ limit
in the exponent. There are several noteworthy points about this
result. The first is that it matches the answer obtained by a
D5 probe brane calculation in $AdS_5\times S^5$
\cite{Yamaguchi:2006tq} and more surprisingly, the D5 brane probe
calculation of the Polyakov loop in the AdS black hole background
\cite{Hartnoll:2006hr}. Although this answer was also obtained by a
matrix model calculation in \cite{Yamaguchi:2006tq}, 
the logic of that calculation is unclear
\footnote{The calculation presented in \cite{Yamaguchi:2006tq} was in fact a 
computation of the cumulative distribution of the first $k$ eigenvalues 
in the cut using the large $N$ distribution 
${2\over\pi}\int_{-1}^{\cos\theta_k} dx x \sqrt{1-x^2}$. Although this 
matches the correct large $\lambda$ result, it differs from the 
correct finite $\lambda$ result as we already see from the 
$O(e^{-\sqrt\lambda})$ corrections in our approach.
Obtaining the correct answer using the approach of 
\cite{Yamaguchi:2006 tq} involves computing the 
cumulative expectation value of 
$k$ distinct eigenvalues sprinkled randomly along the distribution 
governed by the Hermitian matrix model. These will give 
appropriate corrections to the result of \cite{Yamaguchi:2006tq}}.
In particular, our result will have corrections involving powers of 
$e^{-\sqrt\lambda}$ in the exponent. Such corrections were also
present in the matrix model computation of \cite{Drukker:2000rr} for
the Wilson loop in the fundamental representation.
Note also that the result is invariant under $\theta_k\rightarrow\pi
-\theta_k$ reflecting the $k\rightarrow N-k$ symmetry for
completely antisymmetric representations. Finally, as noted in
\cite{Hartnoll:2006hr} and \cite{Yamaguchi:2006tq}, the above form of
the action implies terms that appear nonanalytic from the point of
view of the $1/N$
expansion. Specifically, one obtains contributions involving 
powers of $\left({k\over N}\right)^{2/3}$ upon expanding the
saddle point exponent in powers of $f=k/N$.

{\underline{\bf Symmetric}}: For the rank $k$ symmetric
representation, the saddle point equation corresponds to choosing the
upper signs in \eqref{saddle}. This choice of sign completely changes 
the nature of solutions. First, by separating the equation
into its real and imaginary parts we may show that there are no saddle
points away from the real axis on the $z$ plane. Recall that in the
symmetric case the branch cut is along the interval $-1<
z<1$.

If we start with $\lambda\ll 1$, then we find a saddle point on the
real $z$ axis with ${\rm Re}\; z <-1$. This saddle point can be
treated analytically and we will do so in the following section. As we
increase $\lambda$, however, we find that the saddle point moves in
the positive direction along the real axis. At a critical value,
$\lambda = \lambda_c$, the saddle point hits the branch point at
${\rm Re}\; z = -1$ and appears to move off into the second sheet.
For a given $f$, from (\ref{saddle}) we have that the critical
point is reached when
\begin{equation}\label{critical}
f = - \frac{2}{\pi} \int_{-1}^1 \frac{dx \sqrt{1-x^2}}{1-
  e^{\sqrt{\lambda_c} (x+1)}} \,.
\end{equation}
Solving this equation we obtain a curve that splits the $(f,\lambda)$
plane into regions where a saddle point exists for negative real $z$
and where it does not. This curve is shown in figure 1.

\begin{figure}[h]
\begin{center}
\epsfig{file=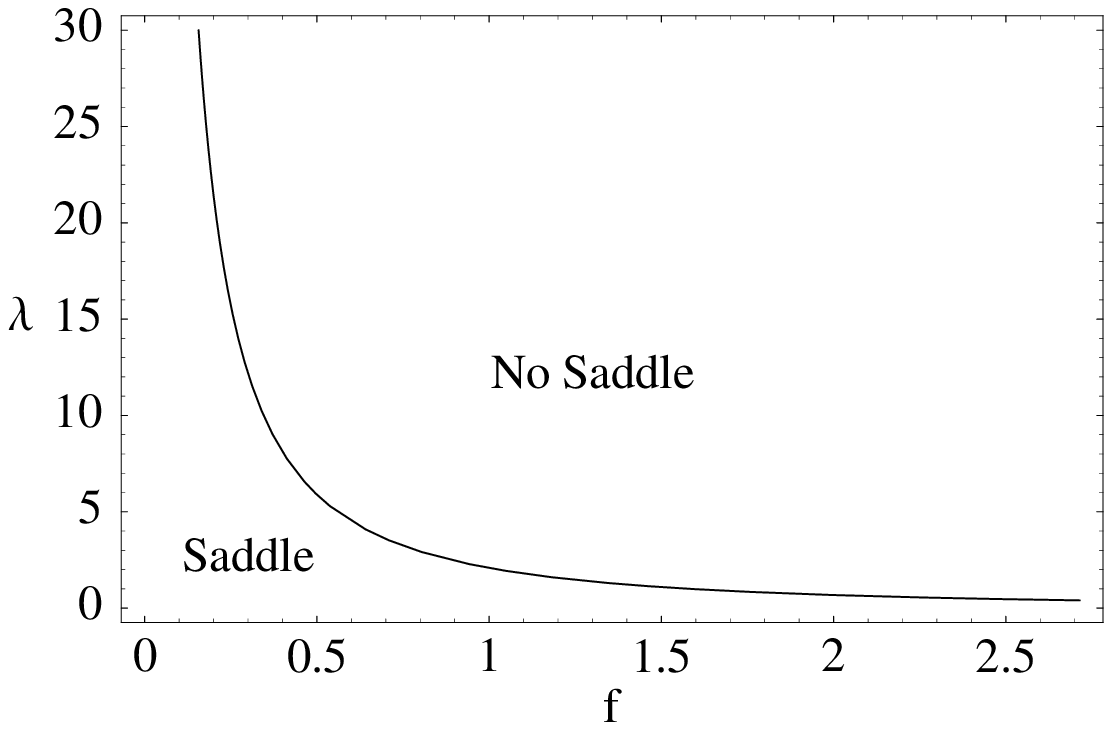,width=8cm}
\end{center}

\noindent {\bf Figure 1:} The existence of a saddle point at negative
real $z$ depends on $f$ and $\lambda$ as shown.

\end{figure}

The asymptotics of the curve in figure 1 may be derived
analytically. We have
\begin{eqnarray}
f \; \simeq \;\frac{2}{\lambda_c^{1/2}} + \cdots
\qquad \text{ as } \qquad \lambda_c \to 0 \,, \nonumber \\
f \; \simeq \; \sqrt{\frac{2}{\pi}} \frac{\zeta(\frac{3}{2})}{
  \lambda_c^{3/4}} + \cdots \qquad
\text{ as }\qquad \lambda_c \to \infty \,.
\end{eqnarray}

In the region where the saddle point exists, we can compute the
location of the saddle point numerically and evaluate (\ref{combin1})
at the saddle, thereby giving us the Wilson loop expectation
value. However, at any fixed $f$, once the coupling is sufficiently
strong there are no saddle points on the first sheet of the complex
plane. We are aiming for analytic expressions in the $\lambda \to \infty$
limit. Since it is not clear how to pick up
saddle point contributions from the second Riemann sheet, we will
adopt a different methodology to evaluate \eqref{combin1} for the
symmetric representation.

The integral has no singularitites at ${\rm Re}\;z\rightarrow+\infty$
(corresponding to $t\rightarrow\infty$ in our initial coordinates).
Its only singularity is the
branch cut on the real axis. We may therefore deform the contour to
wrap the branch cut so that the integral reduces to the
discontinuity across the cut
\SP{\langle\Tr_{S_k}[U]\rangle={{\sqrt\lambda}\over\pi}{\rm Im}\int_{-1}^{1}
  dy\;
\exp\left[- N \left(
    {2\over\pi}  \int_{-1}^{y}
dx\;{\sqrt{1-x^2}}\log(e^{{\sqrt\lambda}y}- e^{{\sqrt\lambda}x})+\right.\right.
\\\left.\left.+ {2\over\pi} \int_{y}^{1}
dx\;{\sqrt{1-x^2}}\log(e^{{\sqrt\lambda}x}- e^{{\sqrt\lambda}y})+
2i\int_{-1}^y dx \sqrt{1-x^2}+
f\sqrt\lambda y \right)\right].\label{disc}}  
The Wilson loop is thus given by a definite integral over the real variable $y$ and we
can unambiguously evaluate it using the stationary phase
method at large $N$, by looking for the saddle points of the integrand analytically
continued to the complex plane.   

We find two sets of saddle points in the large $\lambda$ limit, one
whose location depends only on the combination $f\sqrt\lambda$ in the
strong coupling limit, and another set whose location is independent
of $\lambda$ in this limit. 

The first of these saddle points occurs along the negative real axis
and solves the equation 
\EQ{\label{dfsaddle}{2\over \pi}{\sqrt\lambda}
\int_{-1}^{+1}{dx \sqrt{1-x^2} e^{{\sqrt\lambda}y}\over{e^{\sqrt\lambda
    y}-e^{\sqrt\lambda x}}}+\;4 i\sqrt{1-y^2}+\;f\;{\sqrt\lambda} =0\,.}
The factor of $4$ in this expression arises for the following
reason. Analytically continuing the exponent of \eqref{disc} into
 the complex $y$ plane and in particular to large negative values of
$y$ yields an extra imaginary contribution from the logarithm which
then doubles the coefficient of the third integral in the exponent.

At large $\lambda$ and a fixed value of $\kappa$, where
\EQ{\kappa={f \sqrt\lambda\over 4}\,,}
we find a solution to the saddle point equation (\ref{dfsaddle}) 
along the negative real axis at
\EQ{\label{dfsaddle2} y= y_1= -\sqrt{1+\kappa^2} + O(e^{-{\sqrt\lambda}(|y_1|-1)}) \,.}
Plugging this back into \eqref{disc} we obtain the contribution to
the Wilson loop from this saddle point as
\EQ{\label{matching} W_1=\exp\left[2N\left(\kappa\sqrt{1+\kappa^2}+
\sinh^{-1}{\kappa}\right)\right]\,,}
which is exactly the result found by Drukker and Fiol
\cite{Drukker:2005kx} using a probe D3 brane in $AdS_5\times
S^5$. We have only kept the leading terms in the large $N$,
large $\lambda$ limit in the exponent. In particular we have ignored phases.
This is only one of the two contributions to the symmetric Wilson loop.

The integrand \eqref{disc} appears to have yet another saddle point at large
$\lambda$ in the complex plane, whose position is independent of
$\lambda$ in this limit. To see this we directly take the large 
$\lambda$ limit of the exponent \eqref{disc}. Once we have taken this limit we can
look for the saddle point. The condition turns
out to be very similar to \eqref{sad}, obtained earlier in the antisymmetric case  
\EQ{{2\over \pi}\int_{-1}^y dx \sqrt{1-x^2}+f=0\,,}
except that $y$ can be anywhere in the complex plane. It is obvious
that there are no real solutions to this equation. This is a key
difference between the symmetric and antisymmetric
representations. The expressions obtained are only superficially
similar. It may be checked numerically that the equation has solutions
in the complex plane. In terms of a {\it complex} angular variable, the complex
saddle point in the $y$ plane 
\EQ{y_2 = \cos\phi_k \in {\mathbb{C}}\,,}
is determined by
\EQ{(1+f)\pi = \phi_k - \sin\phi_k\cos\phi_k.\label{ssol}}
This does not have any solutions for real $\phi_k$ between $0$ and
$\pi$. The main difference from the antisymmetric case \eqref{asol} is
that now the left hand side is greater than unity. 

This solution contributes to the 
symmetric Wilson loop in the limit of large
$\lambda$ as 
\EQ{W_2=\exp\left[-{2N\over 3\pi}{\sqrt\lambda}\;{\rm Re} (\sin^3\phi_k)\right]\,,}
where we have ignored phases and terms that are subleading in the
exponent in the large $\lambda$ regime. We have checked numerically
that for generic values of $\lambda$ and $f$ the exponent turns out to
be positive at the saddle point and hence is nonzero in the large $N$ limit.

The Wilson loop in the symmetric representation is given by the sum of
these two stationary point values
\EQ{W_{S_k}[U] = {1\over {\rm dim}[S_k]}\langle\Tr_{S_k}[U]\rangle= W_1+W_2\,.}
Depending on the specific values of $f=k/N$ and $\lambda$, 
one of the two stationary points provides the answer in the large $N$
theory. The limit that is relevant for direct comparison with 
the D3 brane computation in
supergravity is $\lambda \to \infty$ with $\kappa$ fixed. In
this limit the saddle point $W_1$ is dominant and therefore our
computations match the D3 brane result of \cite{Drukker:2005kx}.
Another limit in which $W_1$ dominates is the large $\lambda$ limit
with $k/N$ fixed,
since its exponent scales as $2N\kappa^2\sim N\lambda$ while the
exponent of $W_2$ only scales as $N\sqrt\lambda$. For general values
of $\lambda$ and $f$, either saddle could dominate.

\subsection{Perturbative limit}

Interesting explicit results may also be obtained in the opposite
limit, $\lambda \to 0$. In this regime, a unified treatment of the
symmetric and antisymmetric cases is possible.

The first step is to solve the saddle point condition
(\ref{saddle}) perturbatively
in $\lambda$. As in the previous subsection, the saddle point must lie
on the real axis: $z = \tilde z \in {\mathbb{R}}$. One finds the
following expansion to the lowest two orders 
\begin{equation}\label{pert}
e^{- \sqrt{\lambda} \tilde z} = \left(\pm 1 + \frac{1}{f} \right) +
\frac{\lambda}{8} \left(\pm 3 + \frac{1+2 f^2}{f} \right) + \cdots \,.
\end{equation}
As elsewhere, the upper choice of signs corresponds to the symmetric
case.

Substituting the expansion (\ref{pert}) into (\ref{combin1}) and
expanding the expression in the exponent, we obtain the following
result for the perturbative Wilson loop to leading order as $\lambda
\to 0$
\begin{equation}\label{pertW}
W_{S_k,A_k}[U] = \exp\left[\frac{N \lambda}{8} \left(f \pm f^2
    \right)\right] \,.
\end{equation}
We can recognise the exponent of this result as the quadratic
Casimir of the symmetric and antisymmetric representations of $SU(N)$
in the large $N$ limit with $f = k/N$ kept fixed. This observation
strongly suggests the following perturbative result for a general
representation
\begin{equation}
W_{\cal R}[U] = \exp\left[\frac{\lambda C_2({\cal R})}{8 N} \right] \,.
\end{equation}
The appearance of the quadratic Casimir is not
surprising and the linear dependence on $\lambda$ is consistent with a
one-loop perturbative gauge theory result. The coefficient is a nontrivial result of the
${\mathcal{N}} = 4$ gauge theory.

A more detailed study at weak coupling in the symmetric
case provides a pleasing consistency check of the strong coupling
results. There exists a saddle point which is the weak
coupling counterpart of the saddle (\ref{dfsaddle2}) that we matched
with the supergravity D3 brane result. This may be seen most easily in
the limit $\lambda \to 0$ with fixed $\kappa \gg 1$. In this case the
saddle point equation (\ref{dfsaddle}) is solved by
\begin{equation}
y = y_1 = - \sqrt{1+\kappa^2} + {\mathcal{O}}(\kappa^{-1}) \,.
\end{equation}
Evaluating the Wilson loop on this saddle point yields precisely the
large $\kappa$ limit of the result (\ref{matching}). Thus although
we might appear to be far from the supergravity regime, we can still
make contact with the D3 brane result of \cite{Drukker:2005kx}.

\section{Conclusions}

In this paper we have used the quadratic Hermitian matrix model of
\cite{Erickson:2000af,Drukker:2000rr} to compute circular higher
rank Wilson loops in $SU(N)$
${\mathcal{N}} = 4$ SYM theory. In particular, we gave an
expression for the rank $k$ symmetric and antisymmetric
representations in terms of a single contour integral. This integral
expression is valid for all values of the 't Hooft coupling and at
large $N$ may be computed in the saddle point approximation.

Taking the strong coupling limit, $\lambda \to \infty$, we reproduced
analytically
both the supergravity D3 brane result of Drukker and Fiol \cite{Drukker:2005kx},
using the symmetric representations, and
the supergravity D5 brane result of Yamaguchi \cite{Yamaguchi:2006tq}, using the
antisymmetric representations. This provides
nontrivial evidence for the Wilson loop dictionary proposed in \cite{Gomis:2006sb}
and of the AdS/CFT correspondence more generally: the match involves
higher genus terms, as $k/N$ was kept fixed in the large $N$
limit. Our results
also clear up confusions in the literature, in which the D3 brane
result was initially matched with a multiply wound loop while the D5
brane result was matched with a matrix model computation whose
validity was not a priori obvious.

Away from the strong coupling limit we find corrections to the
supergravity results governed by $e^{-\sqrt{\lambda}}$. These
corrections are computable, if unilluminating at present,
within our framework. We also computed the Wilson loops in the small
$\lambda$ limit, and were led to a simple formula that we expect to be
true for an arbitrary tensor representation with mixed symmetries.

The interpolation from weak to strong coupling appears to be
uneventful in the antisymmetric case. In contrast,
there is a very interesting phenomenon in the symmetric case.
At a critical value of the coupling,
the saddle point of the contour integral hits a branch point and
moves off the real axis and into a second Riemann sheet. We have
shown the dependence of the critical coupling on $k/N$, but it seems
likely that there is interesting physics around this point still
to be uncovered.

The formalism we have presented can also be used at finite $N$, with the
matrix model expectation values evaluated using orthogonal polynomials
rather than contour integrals. This may be another interesting direction
for future work.

\section*{Acknowledgements}

We are particularly grateful to Asad Naqvi for numerous helpful
comments in the course of this work. We would also like to thank Adi
Armoni, Tim Hollowood and Harald Ita. SAH is supported by a
research fellowship from Clare college, Cambridge. SPK is
supported by a PPARC advanced fellowship.

\end{document}